\documentclass[useAMS, referee]{mn2e}
\input{psfig.sty}
\usepackage{graphicx}
\usepackage{times}

\begin{document}
\title{Relation between the hardness ratio and time in the first 2 seconds
for compatible samples of short and long gamma-ray bursts\footnote{send
offprint request to: YunMing Dong}}

\date{2004 Sept. 8}
\pubyear{????} \volume{????}

\pagerange{2} \onecolumn

\author[Y. P. Qin et al.]
       {Yi-Ping Qin$^{1,3}$, Yun-Ming Dong$^{1,2}$
 \\
        $^1$ National Astronomical Observatories/Yunnan Observatory,
         Chinese Academy of Sciences, P. O. Box
110, Kunming 650011, China; \\dongym@ynao.ac.cn\\
        $^2$ Graduate School of The Chinese Academy of
Sciences\\
       $^3$ Physics Department, Guangxi University, Nanning, Guangxi 530004,
P. R. China \\}

\date{Accepted ????. Received ????}

\pagerange{\pageref{firstpage}--\pageref{lastpage}} \pubyear{2004}

\maketitle

\label{firstpage}

\begin{abstract}
In this paper, we randomly selected a sample (sample 1) of long
gamma-ray bursts, with its size being the same as that of the
short burst group (sample 2) (N=500), from the Current BATSE
Bursts Catalog. We randomly selected a short burst and assigned
its T90 to each of these long bursts. We thus constructed a long
burst sample with both the sample size and the distribution of T90
being the same as those of the short burst sample obtained from
the Current BATSE Bursts Catalog. Then we calculated the hardness
ratio ($hr_{T}$) over the assigned T90 for the long bursts and
over their own T90 for the shout bursts, and studied the relation
between the hardness ratio and the corresponding T90 for these two
samples. We also calculated the hardness ratio ($hr_{t}$) over the
randomly selected 64 ms time intervals within the T90, and
investigated the relation between this hardness ratio and the
selected 64 ms time interval. In addition, the $hr_{t}$ within and
beyond the first 2 seconds for all the long bursts (sample 3;
N=1541) were also investigated. We found that the KS probabilities
of the distributions of the $hr_{T}$ (7.15337E-15) and $hr_{t}$
(9.54833E-10) for samples 1 and 2 are very small, and the average
value of $hr_{T}$ and $hr_{t}$ of short bursts are obviously
larger than that of the long bursts. The correlations between
log$hr_{T}$ and logT90, and between log$hr_{t}$ and log t, for
samples 1 and 2 are different. These show that short and long
bursts in the first 2 seconds have different characters and they
probably originate from different progenitors. For sample 3, for
the two time intervals, the KS probability is 5.35828E-5, which
suggests that the hardness ratios in different time intervals for
long bursts are also different.

\end{abstract}

\begin{keywords}
gamma rays:bursts-- methods:statistical
\end{keywords}
\section{Introduction}

Gamma-ray bursts (GRBs) are the brightest astronomical phenomena
since they were firstly detected in the late 1960's (klebesadel et
al, 1973). They almost remained thereafter mysteries for more than
thirties years, largely due to the fact that during this period
they remained detectable only for the high energy--gamma-ray
energies and with short durations. In 1997, the first x--ray
afterglow (GRB970228) (Costa, Frontera \& Heise et al. 1997),
optical afterglow (van Paradijs, Groot \& Galama et al, 1997) and
radio afterglow (GRB 970508)(Frail, Kulkarni \& Nicastro et al.
1997) were detected, which led the study of GRBs into GRB
afterglow era. The discoveries of the GRB afterglows proved that
the GRBs were at cosmological distances. So far, Over forty GRB
afterglows have been detected
(http://www.mpe.mpg.de/$\sim$jcg/grb.html). But all the detected
afterglows belong to the long burst class and there is not any
detected afterglow which is a short burst. This probably indicates
that the two classes of GRBs are intrinsically distinct.
Statistical studies revealed that these two classes have different
distributions of the hardness ratio (short bursts being harder),
pulse width, separation time scale, number of pulses per bursts,
different anti--correlations between the spectral hardness and
duration. It suggests that the two classes might be intrinsically
different (Hurley et al. 1992; Kouveliotou et al. 1993; Fishman \&
Meegan 1995; Norris et al. 2000; Qin et al. 2000, 2001; Nakar \&
Piran 2002). A generally accepted scenario is that short bursts
are likely to be produced by the merger of compact objects while
the core collapse of massive stars is likely to give rise to long
bursts (see, e.g., Zhang \& M$\acute{e}$saz$\acute{a}$ros, 2003;
Piran, 2004).

Recently, Ghirlanda, Ghisellini \& Celotti (2004) have shown that
the emission properties of short bursts are similar to that of the
first 2 seconds for long events and concluded that the central
engine of long and short bursts is the same, only working for a
longer time for long GRBs. Based on this work, Yamazaki, Ioka \&
Nakamura (2004) proposed a unified model of short and long GRBs
and suggested that the jet of GRB consists of multiple subjets or
subshells, where the multiplicity of the subjets along the line of
sight $n_{s}$ is an important parameter. They showed that if
$n_{s}$ is large ($\gg$ 1), the event looks like a long GRB, while
if $n_{s}$ is small ($\sim$ 1), the event looks like a short GRB.
Based on the unified model of Yamazaki et al. (2004), Toma,
Yamazaki \& Nakamura (2004) have successfully explained the
bimodal distribution of T90 of the GRBs. Thus, it is still unclear
whether the short and long bursts are intrinsically the same. In
this paper, we will adopt the definition of the hardness ratio
associated with time, ($hr_{t}$), presented in Dong \& Qin (2004),
and pay our main attention to the differences of the quantity
within the first 2 seconds for these two groups of GRBs. Owing to
the observed event rate of short bursts being $\sim$ 1/3 of that
of the long bursts, we randomly selected a sample from the whole
long bursts, with its number being the same as that of the short
burst group.

This paper is organized as follows. In section 2, we described a
long burst sample with both the sample size and the distribution
of T90 being the same as those of the short burst sample obtained
from the Current BATSE Bursts Catalog. Then we calculated the
hardness ratio ($hr_{T}$) over the assigned T90 for the long
bursts and over their own T90 for the shout bursts, and studied
the relation between the hardness ratio and the corresponding T90
for these two samples. We also calculated the hardness ratio
($hr_{t}$) over the randomly selected 64 ms time intervals within
the T90, and investigated the relation between this hardness ratio
and the selected 64 ms time interval for these two samples. In
section 3, we investigated the $hr_{t}$ in different time
intervals for all the long bursts. A brief discussion was
presented in section 4.

\section{Relation between the Hardness Ratio and Time for Short and Randomly Selected Long Bursts}

In the duration table of the Current BATSE Bursts Catalog
(http://cossc.gsfc.nasa.gov/batse/), 1541 long bursts and 500
short bursts are presented. The 64 ms temporal resolution and
four-channel spectral resolution GRB data (Concatenated 64-ms
Burst Data in ASCII Format) observed by BATSE can also be
available via anonymous ftp in the web site
(ftp://cossc.gsfc.nasa.gov/compton/data/batse/). The data type of
Concatenated 64-ms Burst Data in ASCII Format is a concatenation
of three standard BATSE data types, DISCLA, PREB, and DISCSC. All
three data types are derived from the on-board data stream of
BATSE's eight Large Area Detectors (LADs), and all three data
types have four energy channels, with approximate channel
boundaries: 25-55 keV, 55-110 keV, 110-320 keV, and $>$320 keV.
These data were used to calculate the hardness ratios in this
paper. Owing to the observed event rate of short bursts is $\sim$
1/3 of that of long bursts, we randomly selected a long bursts
sample (sample 1), with its size being the same as that of the
short burst group (sample 2, N=500), from all the long bursts
(sample 3, N=1541) available in the duration table. But when
performing the calculation of the hardness ratio in this paper, we
found that some bursts presented in the duration table do not have
the corresponding 64 ms temporal resolution and four-channel
spectral resolution GRBs data in the Concatenated 64-ms Burst Data
in ASCII Format. Thus, in this paper, only 462 short bursts and
467 long bursts were employed to calculate the following hardness
ratio.

The durations of long bursts are larger than 2 seconds, while they
are less than 2 seconds for short bursts. In order to investigate
the hardness ratio of the two GRBs classes within the first 2
seconds, we considered two definitions of the hardness ratio.
First, we assigned a duration T90 of a randomly selected short
burst from sample 2 to each of the long bursts in sample 1, and
calculated the hardness ratio with the sum of the counts within
this assigned T90, ($hr_{T}$), for each long burst, and calculated
the hardness ratio with the sum of the counts within its own T90
for every short burst. Second, the hardness ratio is calculated
with the count in a randomly selected 64 ms time interval,
($hr_{t}$), within the assign T90 for long bursts and within their
own T90 for short bursts. The time intervals defining $hr_{T}$ is
from the beginning of the T90 to the end of the T90 considered
here. The $hr_{T}$ is determined by
\begin{equation}
hr_{T}=\frac{\sum\limits_{0}^{T90}Count3}{\sum\limits_{0}^{T90}Count2}
\end{equation}

where count3 is the count of the third channel of the Concatenated
64-ms Burst Data in each 64 ms time intervals, and count2 is that
of the second channel.

The plot of the log$hr_{T}$--logT90 for the short and selected
long bursts is presented in Fig. 1. In this plot, all data points
are presented and the regression lines for the two classes are
drawn. Presented in the plot are also two data points standing for
the average values of the two quantities for the two classes. The
corresponding distributions of the $hr_{T}$ of the two classes are
shown in Fig. 2, and the probability of KS test to the two
distributions is 7.15337E-15. We find : (1) the correlation
coefficient between the two quantities for short bursts is
r=-0.19, where the size of the short bursts is N=462; and for the
selected long bursts it is r=-0.15, where the size of the long
bursts is N=467. This shows that for the two classes the
log$hr_{T}$ and the logT90 are correlated, but the correlation of
the two classes is different. As the KS probability is very small,
which is only 7.15337E-15, it shows that the distributions of the
$hr_{T}$ of the two GRBs classes are obviously different,
indicating that they are likely to arise from different
distributions. Meanwhile, we also find in Fig. 1 the average value
of the log $hr_{T}$ for the short bursts (0.0146) is larger than
that of the long bursts (-0.037), which is consistent with the
previous results (Dezalay, Lestrade \& Barat et al. 1996).

\begin{figure}

  \includegraphics[width=6in,angle=0]{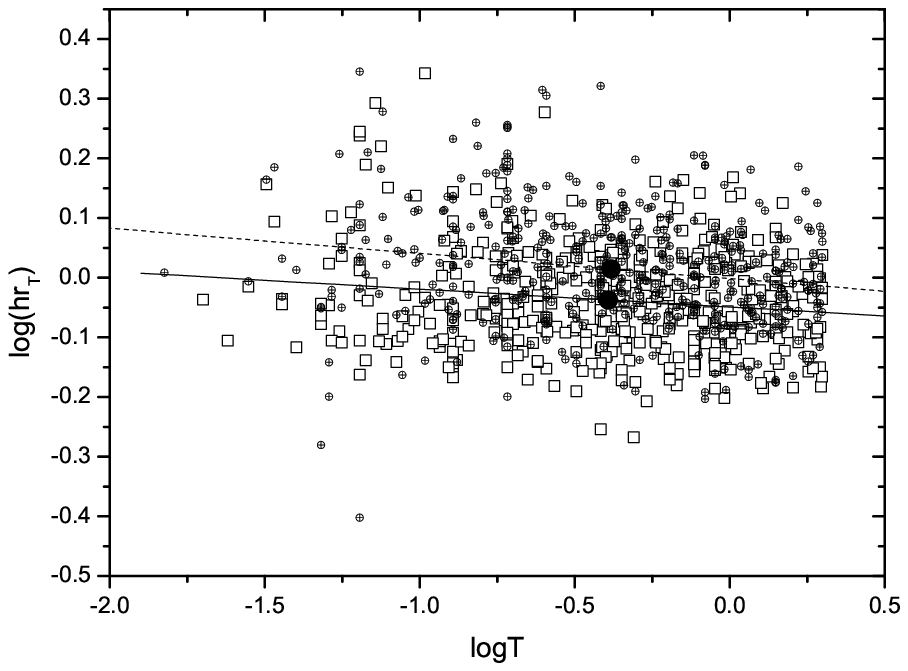}

   \caption{The plot of log$hr_{T}$--logT90 for the short bursts and selected
   long bursts. The open squares represent the long bursts, and the
   open circles plus cross represent the short bursts,
   respectively. The dashed line is the regression line for the
   short bursts and the solid line is the regression line for the
   selected long bursts. The solid circles represent the two data
   points standing for the average values of the two quantities
   for the two classes, respectively.}
  \label{Fig1}
\end{figure}

\begin{figure}

  \includegraphics[width=6in,angle=0]{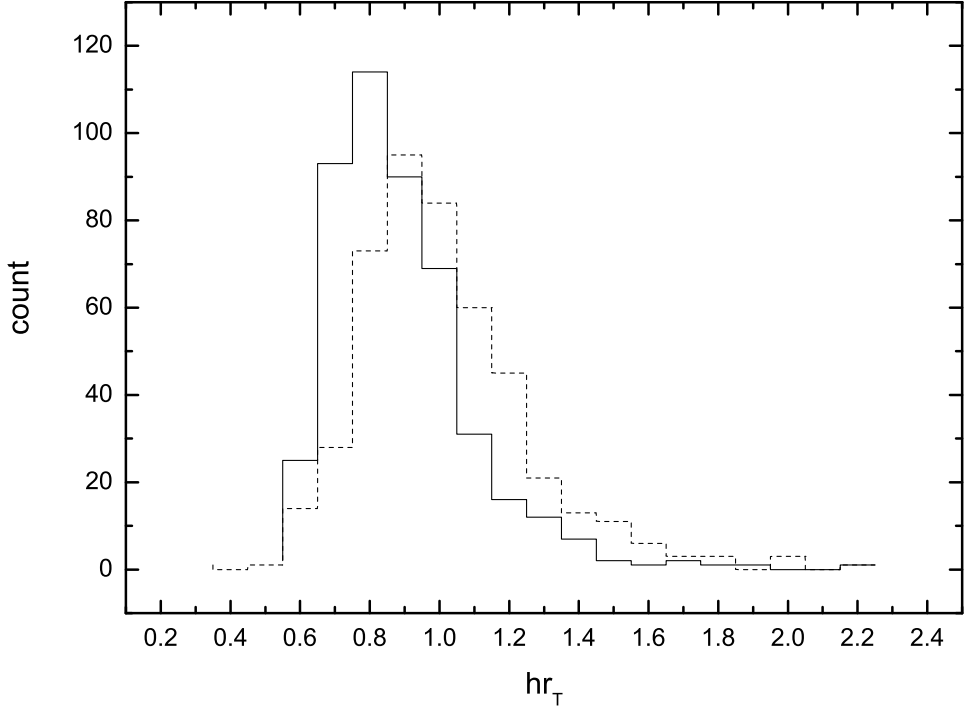}

   \caption{The distributions of $hr_{T}$ of the short bursts and selected
   long bursts. The dashed line represent the
   short bursts and the solid line represent the
   selected long bursts. The probability of the KS test to the two distributions is
   7.15337E-15}
  \label{Fig2}
\end{figure}

In the second situation, the way of calculating the hardness ratio
over a randomly selected 64 ms time intervals for short and these
long bursts is as follows. We firstly randomly, and not
repeatedly, selected a short and long burst, then randomly
selected a 64 ms time intervals within the T90 of this short
burst. Secondly, we calculated the hardness ratio with the count
in this 64 ms time interval for this short and long bursts. Thus,
for any of short and long bursts, we can calculate their hardness
ratio associated with the randomly selected 64 ms time interval.
The $hr_{t}$ is determined by
\begin{equation}
hr_{t}=count3/count2
\end{equation}

where count3 is the count of the third channel of the Concatenated
64-ms Burst Data in the randomly selected time intervals, count2
is that of the second channel, and t is the corresponding time of
the selected time interval measured from the beginning of the T90.

The plot of the log$hr_{t}$--log t for the short and selected long
bursts is shown in Fig. 3. In this plot, all data points are
presented and the regression lines for the two classes are drawn.
Presented in the plot are also two data points standing for the
average values of the two quantities for the two classes. At the
same time, the distributions of $hr_{t}$ for the selected long
bursts and short bursts are presented in Fig. 4, and the
probability of the KS test to the two distributions is
9.54833E-10, which is also very small. In Fig. 3, we can find: the
correlation coefficient between the two quantities for the short
bursts is r=-0.17, and for the long bursts it is r=0.046. They are
correlated for the short bursts but not correlated for the long
bursts. The average value of the log$hr_{t}$ for the short bursts
(0.0049) is larger than that of the long bursts (-0.039). Thus,
from these two situations we can find that short bursts are
obviously different from long bursts. This shows that the two GRB
classes should be intrinsically distinct, and they should
originate from different progenitors.

\begin{figure}

  \includegraphics[width=6in,angle=0]{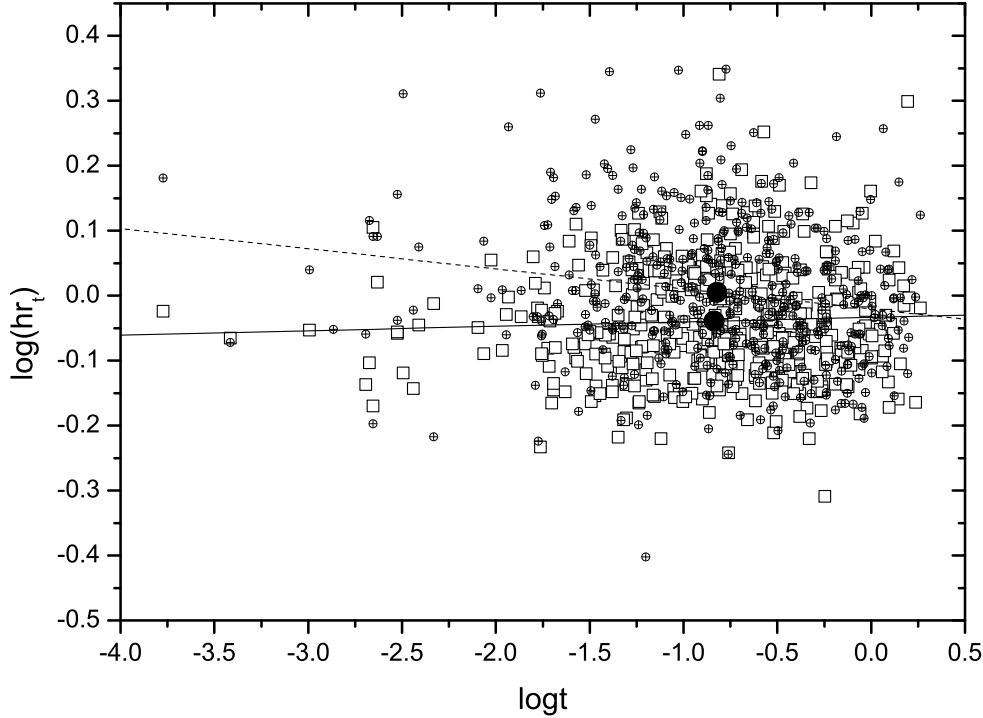}

   \caption{The plot of log$hr_{t}$--log t for the short bursts and the selected
   long bursts. The open squares represent the long bursts, and the
   open circles plus cross represent the short bursts,
   respectively. The dashed line is the regression line for the
   short bursts and the solid line is the regression line  for the
   selected long bursts. The solid circles represent the two data
   points standing for the average values of the two quantities
   for the two classes, respectively.}
  \label{Fig3}
\end{figure}

\begin{figure}

 \vspace{5.5cm}
  \includegraphics[width=6in,angle=0]{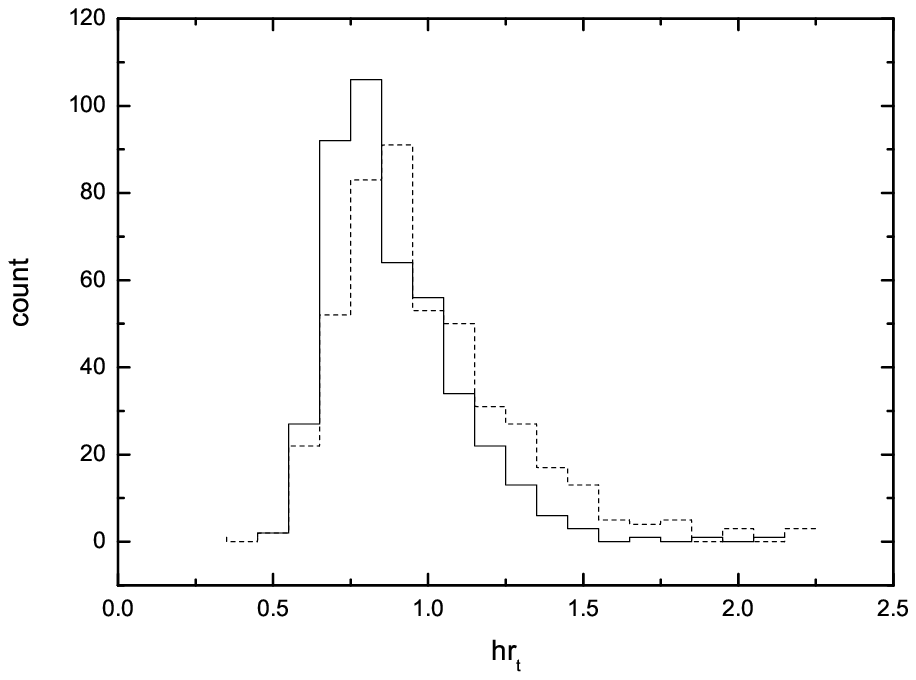}

   \caption{The distributions $hr_{t}$ for the short and selected
   long bursts. The dashed line represent the
   short bursts and the solid line is the regression line represent the
   selected long bursts. The probability of the KS test to the two distributions is
     9.54833E-10}
  \label{Fig4}
\end{figure}

\section{Relation between the Hardness Ratio and Time within and beyond the First 2 Seconds for the Long Bursts}

Ghirland et al. (2004) suggested that the central engine of short
and long bursts is the same, just working for a longer time for
long bursts and they showed that the emission properties of short
bursts are similar to that of the first 2 seconds of long bursts.
Thus, in section 2, we investigated the $hr_{T}$ and the $hr_{t}$
for short and randomly selected long bursts within the first 2
seconds, and found that they are obviously different, indicating
that they should not arise from a same physical process. In this
section, we will investigate the hardness ratio for all the long
bursts (sample 3) in different time intervals. We divided the
duration of a long burst into two time intervals. The first time
interval is from the beginning of T90 to the first 2 seconds, and
the second time interval is from the first 2 seconds to the end of
T90. In each time interval, we randomly selected a 64 ms time
interval and calculated the corresponding hardness ratio $hr_{t}$
with the count in this 64 ms time interval. We therefore obtained
two set of $hr_{t}$ for long bursts. The two set of $hr_{t}$ will
be used to investigate whether there are important differences
between the two time intervals for long bursts.

The plot of the log$hr_{t}$ --log t in the two different time
intervals for long bursts is shown in Fig. 5a, and the plot of the
log$hr_{t}$ --log t for short bursts (sample 2) within the first 2
seconds and for long bursts (sample 3) beyond the first 2 seconds
is shown in Fig. 5b. In Fig. 5a, all data points are presented and
the regression lines for the two time intervals are drawn.
Presented in Fig. 5a are also two data points standing for the
average values of the two quantities for the two time intervals.
In Fig. 5b, the symbols are the same as those in Fig. 5a, but they
represent two groups of GRBs (samples 2 and 3). In Fig. 6 we
present the distributions of $hr_{t}$ within and beyond the first
2 seconds for the long bursts. The probability of the KS test to
the two distributions is 5.35828E-5. This probability is so small
that we can not think they are the similar distributions for the
two time intervals, which shows that the hardness ratios in
different time intervals for long bursts are different. From Fig.
5a we can find: the correlation coefficient between the two
quantities in the first 2 seconds is r=-0.0286 and it is r=0.0237
beyond the first 2 seconds for the long bursts. This shows that
for the two time intervals the log$hr_{t}$ and log t are not
correlated. We also find that the regression lines of the two
quantities within and beyond the first 2 seconds are almost
parallel, and the average value of the log$hr_{t}$ in the first 2
seconds (-0.043) is appreciably larger than that beyond the first
2 seconds (-0.061). But from Fig. 5b, we find that the difference
of the log $hr_{t}$--log t correlation between the short and long
bursts is obvious. Firstly, the average value of log$hr_{t}$ of
short bursts (0.0049) is obviously larger than that of the long
bursts (-0.061) beyond the first 2 seconds. Secondly, the
regression lines of the short within the first 2 seconds and of
the long bursts beyond the first 2 seconds are obviously
different. This shows clearly that the difference between the two
classes of bursts is not at all due to time interval (if due to
time interval, the two regression lines within the first 2 seconds
in Figs 5a and 5b should be almost the same). The two class are
likely to be distinct groups.
\begin{figure}

  \includegraphics[width=6in,angle=0]{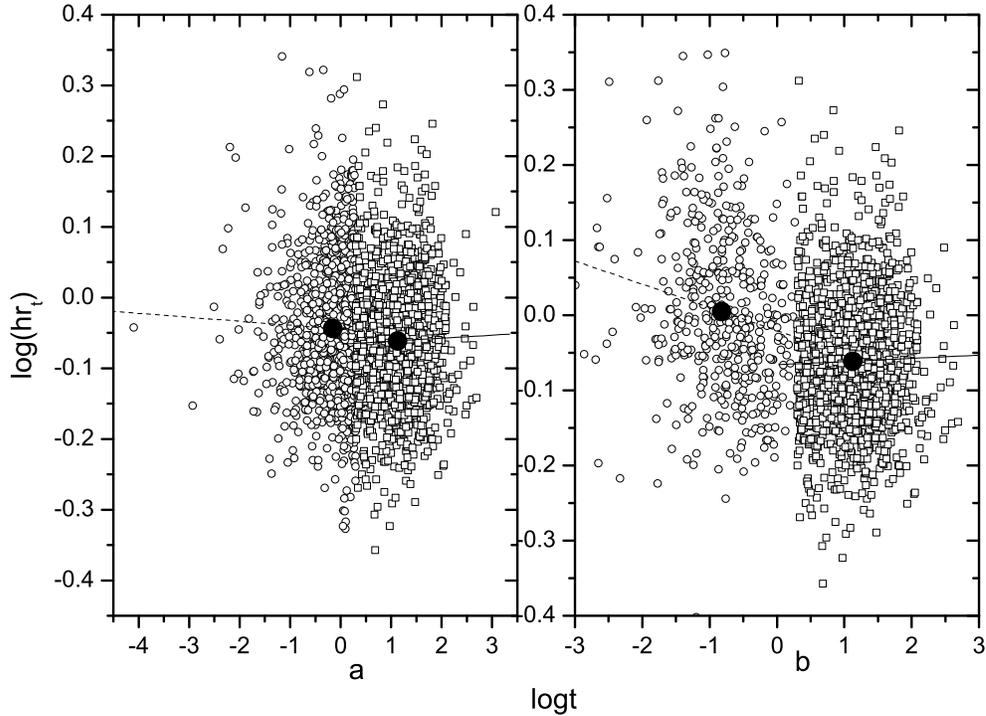}

   \caption{a: The plot of log$hr_{t}$--log t for long bursts within (open circles) and beyond (open squares) the first 2
   seconds, respectively. The dashed line is
   the regression line for the data within the first 2 seconds and the solid line is the regression line  for
   that beyond the first 2 seconds. The solid circles represent the two data
   points standing for the average values of the two quantities
   for the two classes respectively. b: The plot of log $hr_{t}$-log t for the data of short bursts (sample 2; open circles)
   within the first 2 seconds and for that of long bursts (sample 3; open squares) beyond the first 2 seconds. The dashed line
   is the regression line for the short bursts and the solid line
   is the regression line for the long bursts. The solid circles represent the two data
   points standing for the average values of the two quantities
   for the two classes, respectively.}
  \label{Fig5}
\end{figure}

\begin{figure}

  \includegraphics[width=6in,angle=0]{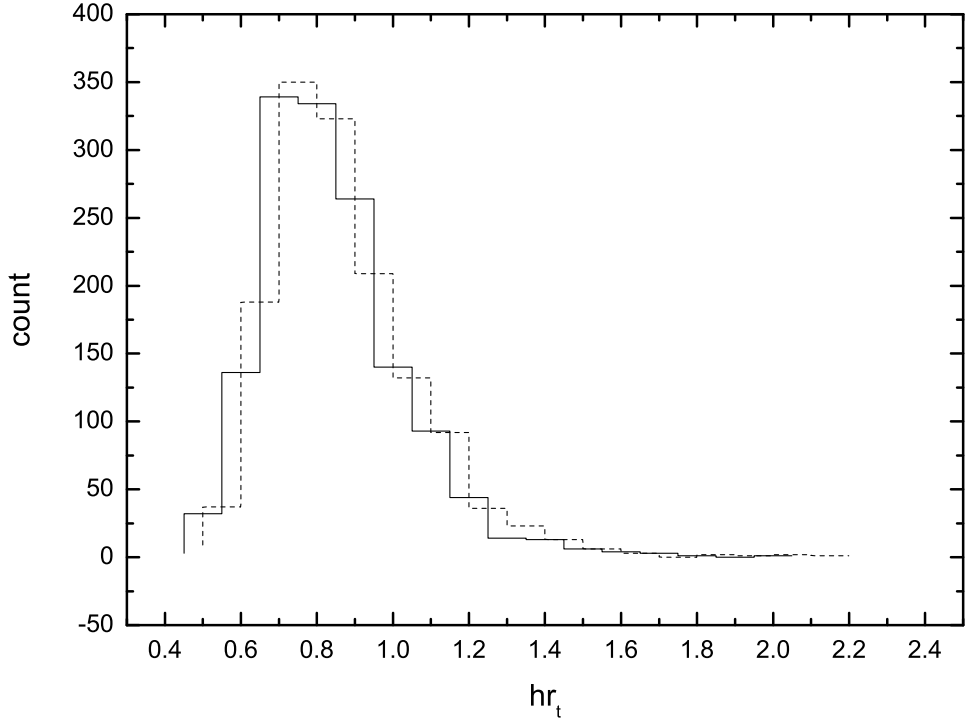}

   \caption{The distributions of $hr_{t}$ for
   long bursts in different time intervals. The dashed line represent the data of long bursts in the
   first 2 seconds and the solid line stands for that of the long bursts beyond
   the first 2 seconds. The possibility of KS test to the two distributions is
   5.35828E-5}
  \label{Fig6}
\end{figure}

\section{Discussions and conclusions}

In this paper, we randomly selected a long bursts sample, with its
size being the same as that of the short burst group, from all the
long bursts available in the Current BATSE Bursts Catalog. We
randomly, and not repeatedly, selected a short burst and assigned
its T90 to one of these long bursts as its new T90. Then we
investigated the hardness ratio ($hr_{T}$) over the assigned T90
for the long bursts and over their own T90 for the short bursts.
Meanwhile, we also investigated the hardness ratio associated with
the randomly selected 64 ms time intervals ($hr_{t}$) within the
T90 of short bursts for both short and long bursts. In addition,
the $hr_{t}$ within and beyond the first 2 seconds for all the
long bursts were also investigated. The main aim of this work is
to find whether the hardness ratio of short and long bursts in the
first 2 seconds possess the same character.

In section 2, one can find for short and randomly selected long
bursts that, the KS probabilities of the distributions of the
$hr_{T}$ (7.15337E-15) and $hr_{t}$ (9.54833E-10) are very small,
and the average value of $hr_{T}$ and $hr_{t}$ of short bursts are
obviously larger than that of the long bursts. The correlations
between log$hr_{T}$ and logT90, and between log$hr_{t}$ and log t,
for short and randomly selected long bursts are different. These
show that short and long bursts in the first 2 seconds have
different characters and they probably originate from different
progenitors. In section 3, for the long bursts, the correlations
between log$hr_{t}$ and log t in different time intervals are
similar and the average values of $hr_{t}$ do not show an obvious
difference. But the ks probability of distributions of the
$hr_{t}$ (5.35828E-5) for long bursts in the two time intervals is
small. These suggest that the hardness ratios in different time
intervals for long bursts are similar in some aspect, but
meanwhile, to a certain extent, they also show some differences.
When replacing the data of long burst with that of short burst
within the first 2 seconds, the situation becomes much different:
the regression line now deviates obviously from that of the long
bursts beyond the first 2 seconds. This suggests that the
difference between long and short bursts is intrinsical and the
two class are likely to be distinct groups. Thus, we can draw
these conclusions: (1) the distributions of the $hr_{T}$ or
$hr_{t}$ of short bursts and the long bursts in the first 2
seconds can not arise from the same parent population and the two
classes probably originate from different progenitors; (2) the
average value of $hr_{T}$ or $hr_{t}$ for short bursts and the
long bursts is different and that of the short bursts is larger
than that of the long bursts, in agreement with previous study;
(3) for the short and long bursts, the correlation between the
hardness ratio and the corresponding time is different; (4) the
hardness ratios in different time intervals for long bursts are
not the same, in consistent with the well-known hard-to-soft
character observed previously.

We would like to address our great thanks to the anonymous referee
for his or her several incisive and substantive comments and
suggestions, which improves this paper greatly. This work was
Supported by the Special Funds for Major State Basic Research
("973") Projects and the National Natural Science Foundation of
China (No. 10273019).

\label{lastpage}

\end{document}